\documentstyle[12pt,a4]{article}

\font\tenrm=cmr10

 1
 1
 1

\newcommand{\rd}{{\mathrm{d}}}
\newcommand{\re}{{\mathrm{e}}}
\newcommand{\rp}{{\mathrm{p}}}
\newcommand{\rB}{{\mathrm{B}}}
\newcommand{\rC}{{\mathrm{C}}}
\newcommand{\seg}{s_{\re \gamma}}
\newcommand{\qqbar}{{\mathrm{q}\bar{\mathrm{q}}}}
\begin{document}
\renewcommand{\thefootnote}{\fnsymbol{footnote}}
\thispagestyle{empty}
\begin{flushright}
CERN-TH/95-233
\end{flushright}
\vskip 2.cm
\begin{center}{RADIATIVE CORRECTIONS TO {\large$\re\gamma$} SCATTERING}
\vglue 2.5cm
\begin{sc}
Eric Laenen$^a$ \footnote{E-mail address: {\tt laenen@surya11.cern.ch}}
and Gerhard A. Schuler$^{a,b}$
\footnote{E-mail address: {\tt schulerg@cernvm.cern.ch}}\\
\vglue 0.6cm
\end{sc}
$^a\,${\it CERN TH-Division\\
1211-CH, Geneva 23, Switzerland}
\vglue 0.2cm
$^b\,${\it Theoretische Physik, Universit\"{a}t Regensburg\\
D-93053 Regensburg, Germany}
\end{center}
\vglue 1cm

\renewcommand{\thefootnote}{\arabic{footnote}}
\setcounter{footnote}{0}

\noindent
\begin{abstract}
We investigate the effects of photon
radiation on deep-inelastic $\re\gamma$ scattering.
Depending on the set of variables chosen, we
find appreciable effects in the kinematic region 
accessible at LEP2. Convenient analytic results
for $O(\alpha)$ corrected differential cross sections
are presented.
\end{abstract}

\vfill
\begin{flushleft}
CERN-TH/95-233\\August 1995
\end{flushleft}

\newpage

\setcounter{page}{1}

\noindent 

\section{Introduction}

In the period leading up to the start of the HERA program
a substantial amount of work was done on radiative corrections
to observables in deep-inelastic scattering (DIS) of electrons
off protons (see e.g. \cite{HERAproc} for an overview). These corrections
are substantial enough for both HERA experiments to correct for them.
Part of the upcoming LEP2 physics programme involves the measurement
of the photon structure function $F_2^{\gamma}(x,Q^2)$.
This structure function is extracted from deep-inelastic electron--photon 
scattering in the reaction $\re^+\re^- \rightarrow \re^+\re^- X$, 
where one of the leptons escapes undetected down the beam pipe, 
while the other is measured at rather large angle.
It is therefore important to study radiative corrections 
to deep-inelastic electron--photon scattering. This is our
purpose in this paper.

Radiative corrections to $\re^+\re^- \rightarrow \re^+\re^- X$ 
have been calculated for $X$ a (pseudo) scalar particle 
\cite{Defrise,Neerven,Landro} and $X = \mu^+\mu^-$ \cite{Landro,Berends}. 
It was found that they are very small 
(on the percent level) in the no-tag case, 
when neither the electron nor the positron is being measured. 
In such a kinematic configuration, when the
momentum transfer between the incident 
and outgoing electron (or positron) is small, the vertex- 
and bremsstrahlung contributions effectively cancel each other, 
leaving a small correction dominated by 
vacuum polarization \cite{Neerven,Landro}. This implies that 
for the equivalent photon spectrum, which is essential to relate 
$\re^{\pm}\gamma$ with $\re^+\re^-$ reactions, corrections are small. 
In contrast, radiative corrections can be sizeable in the case 
where one of the leptons scatters
at a large angle - single tag - and one 
studies differential cross sections 
which depend strongly on the energy and angle of the tagged 
electron (or positron). 

Surprisingly, no calculations of the size of radiative corrections 
for inclusive deep-inelastic electron--photon scattering, i.e. 
$e\gamma \rightarrow e X$ with both large $Q^2$ (the
absolute value of the transferred momentum squared) and $W$ (the mass
of state $X$), have been performed yet. Correspondingly, 
in experimental analyses of the (hadronic) photon structure function 
$F_2^\gamma(x,Q^2)$ radiative corrections have so far not been included
(as usual, $x=Q^2/(Q^2+W^2)$).
They are usually 
assumed to be negligible. Recently, the AMY collaboration \cite{AMY}
estimated the size of radiative corrections by comparing 
a Monte-Carlo event generator based on the full cross section formula
for $\re^+\re^- \rightarrow \re^+ \re^-  \gamma\mu^+\mu^-$ \cite{Berends} 
(with the muon mass and electric charge changed to correspond
with quark-antiquark pair production) 
with a generator for $\re^+\re^- \rightarrow \re^+\re^- \qqbar$ where 
the cross section for $\re \gamma \rightarrow \re \qqbar$ with 
a real photon was convoluted with the equivalent photon spectrum.  
A (positive) correction  
of order $10$\% for the visible $x$ distribution was found, which, 
however, cancelled effectively against the correction due to a non-vanishing 
target-photon mass. Hence no net correction was applied.

Nevertheless, it is important to understand both corrections separately,
and their behavior as a function of the kinematic variables chosen
and phase space. Moreover, the hadronic structure of the photon
must not be neglected.
Here we therefore estimate the size of the radiative corrections 
to inclusive deep-inelastic $\re\gamma$ scattering, using the full
photon structure function.

In the next section we describe the relevant kinematics and formalism
and in section 3 we present results.

\section{Formalism}

We consider the $O(\alpha)$ corrections to deep-inelastic 
scattering (DIS) of electrons on (quasi-real) photons:
\begin{equation}
 \re(l) + \gamma(p) \rightarrow \re(l') + \gamma(k) + X(p_X) 
\ .
\label{DISreact}
\end{equation}
This process is depicted in Fig.1, in which we indicate
all momentum labels.

The target photon $\gamma(p)$ is part of the flux of equivalent
photons around the non-tagged lepton. We assume that 
this flux has a momentum density given 
by the Weizs\"{a}cker-Williams expression
\begin{equation}
  f_{\gamma/\re}(z) = \frac{\alpha}{2\pi} \left\{
  \frac{1+(1-z)^2}{z}\, \ln \frac{P^2_{max}}{P^2_{min}}
  - 2 m_e^2 z(\frac{1}{P^2_{min}}
 -\frac{1}{P^2_{max}})\right\}
\label{fww}
\end{equation}
where $P^2_{min}  = (z^2 m_e^2)/(1-z)$ and $P^2_{max}  = 
 (1-z) \left(E_b \theta_{max}\right)^2$.
Here $z$ is the longitudinal momentum fraction of the 
target photon with respect to its parent lepton, $E_b=\sqrt{s}/2$ 
is the lepton beam energy, $\theta_{max}$ is the anti-tag
\footnote{i.e. all events in which the parent lepton scatters 
at an angle larger than $\theta_{max}$ are rejected.} angle
and $P^2 = -p^2$.
In the following we put $P^2 = 0$ and neglect electron masses
everywhere except in (\ref{fww}). Moreover we substitute
$P^2_{max}$ by $P^2_{max}+P^2_{min}$ so that we can easily
extend the $z$ range to 1, see \cite{FMNR}.

\vglue 6.1cm

\vbox{\includegraphics{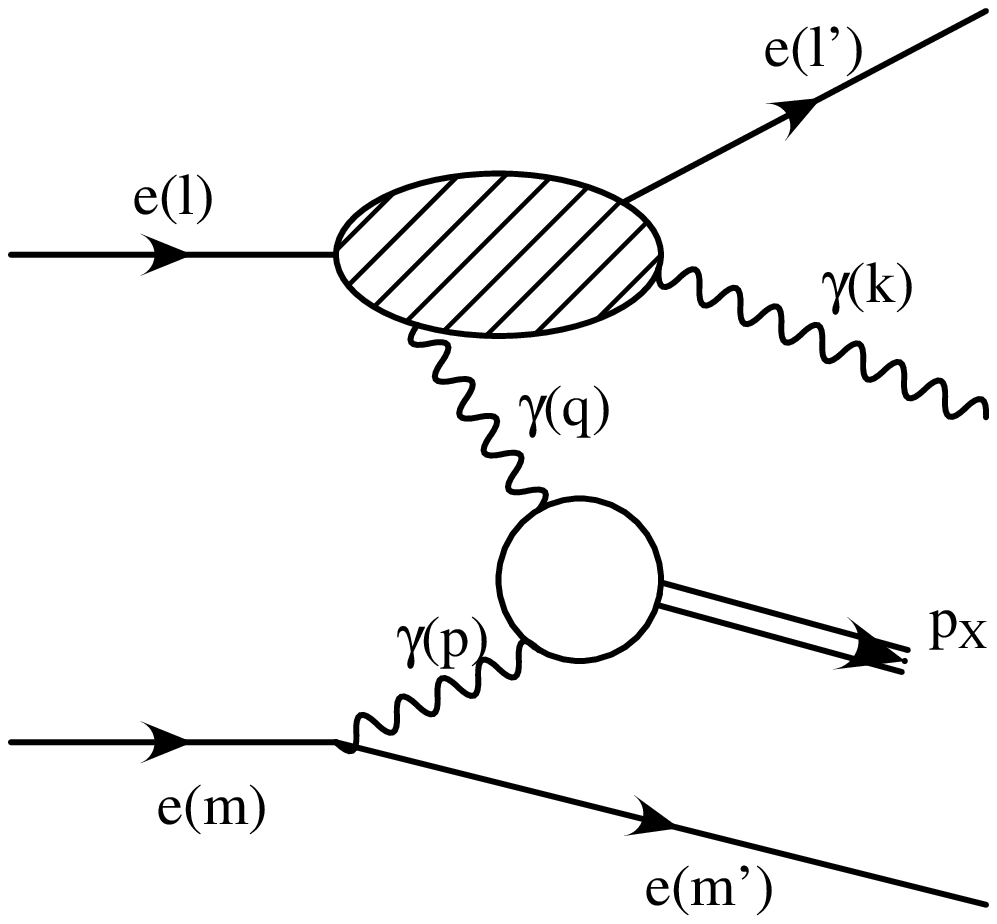}}      

{\tenrm\baselineskip=12pt 
\noindent Figure 1: Photon bremsstrahlung from the tagged lepton line
in deep-inelastic scattering off an equivalent photon.}
\vglue 0.8cm

The DIS variables can be defined either from the leptonic or 
the hadronic momenta:
\begin{equation} 
\begin{array}{rclrcl}
 q_l & = & l-l' & q_h & = & p_X - p = q_l - k\\
 W_l^2 & = & (p+q_l)^2 \qquad & W_h^2 & = & (p+q_h)^2 = p_X^2\\
 Q_l^2 & = & - q_l^2 &  Q_h^2 & = & - q_h^2\\
 x_l & = & Q_l^2/2p \cdot q_l & 
     x_h & = & Q_h^2/2p \cdot q_h\\
 y_l & = & p \cdot q_l/p \cdot l & 
 y_h & = & p \cdot q_h/p \cdot l
\end{array} 
\label{DISvar}
\end{equation} 
Note that both $Q_l^2 = x_l y_l \seg$ and  $Q_h^2 = x_h y_h \seg$, 
where $\seg = (p+l)^2$,
but that leptonic and hadronic variables agree only for nonradiative
events, i.e. if $k=0$. 
We will see that the size of the corrections strongly
depends on which set of variables are used 
in the measurement. In practice one determines $Q^2$ from
the tagged lepton, and $x$ from the (visible) hadronic
energy $W_h$.

The Born cross section (i.e. no $\gamma(k)$ in (\ref{DISreact}))
is given by 
\begin{equation}
  \frac{\rd^2\sigma^\rB}{\rd x \rd Q^2} = f^\rB(x,Q^2,s) 
\ ,
\label{sigBorn}
\end{equation}
where
\begin{eqnarray}
  f^\rB(x,Q^2,s) &= &   \frac{2\pi\alpha^2}{x Q^4}\, F_2(x,Q^2) 
  \\ & \times &  \int_{z_{min}(x,Q^2,s)}^1 \rd z\, f_{\gamma/\re}(z)\, 
      Y_+\left(\frac{Q^2}{x z s} \right) \, 
  \left\{ 1 + R\left(x,Q^2,\frac{Q^2}{x z s} \right) \right\} 
\label{Borncross} \ .
\end{eqnarray}
Here we have defined $z_{min}(x,Q^2,s)  =  Q^2/xs$, 
$ Y_+(y)  =  1 + (1-y)^2$ and 
\begin{equation}
 R(x,Q^2,y)  =   \frac{- y^2}{1 +(1-y)^2} \ 
            \frac{F_L(x,Q^2)}{F_2(x,Q^2)} \ .
\label{Rdef}
\end{equation}
$F_{2,L}$ are the photon structure functions (we have dropped 
the superscript $\gamma$).
The above form eq.~(\ref{Borncross}) is useful because $F_2$ can be factored 
out of the $z$ integration. 
For comparison of $\re \gamma$ with $\re \rp$ scattering we will also 
give the cross section in terms of $x$ and $y$
\begin{equation}
  \frac{\rd^2\sigma^\rB}{\rd x \rd y} = g^\rB(x,y,s) 
\ ,
\label{nsigBorn}
\end{equation}
where
\begin{equation}
 g^\rB(x,y,s)  = \frac{2 \pi\alpha^2 \,Y_+(y)}{x^2 y^2 s} 
   \int_{\epsilon(x,y,s)}^1 \frac{\rd z}{z} \,  f_{\gamma/\re}(z)\, 
   F_2(x,xyzs) \left\{1 + R(x,xyzs,y) \right\}
\ .
\label{newBorn}
\end{equation}
The lower limit $\epsilon = W^2_{\mathrm{min}}/(1-x)ys$
on the $z$-integration in eq.~(\ref{newBorn}) arises if a lower 
cut is applied to the invariant hadronic mass $W$.
At the Born level expressions (\ref{sigBorn}) and (\ref{nsigBorn})
are equivalent and valid for both sets of variables in (\ref{DISvar}).
For the $\re$p case the full electroweak 
corrections have been calculated for both neutral current reactions
\cite{Dubna,Wurzburg},
including elastic nucleon scattering \cite{Akhundov},
and charged-current reactions \cite{Bohm}. 
It is well-known that the leading logarithmic 
approximation (LLA) (in $\ln(Q^2/m_e^2)$) reproduces the 
exact results to within a few percent however
\cite{Consoli, Spiesberger}.
Here we will therefore work within this approximation
and neglect furthermore $Z$ exchange.

We consider first the radiative corrections to the 
differential cross section in (\ref{nsigBorn})
in terms of $x_l$ and $y_l$.
The $O(\alpha)$ corrections in LLA arise from 
collinear and soft bremsstrahlung from 
the initial and final electron that couple to the ``probing photon" 
and from the Compton process\footnote{As stated above,
radiative effects to the Weizs\"{a}cker-Williams spectrum
are small \cite{Landro} and we therefore neglect them.}. 
The latter corresponds to the case
in Fig.1 where the exchanged photon $\gamma(q)$ is quite
soft, but $\gamma(k)$ is radiated at a wide angle
causing the lepton $e(l')$ to be tagged.
The bremsstrahlung terms are given by 
\begin{equation}
 \frac{\rd^2\sigma^{\mathrm{Br}}}{\rd x \rd y} = 
   \int_0^1 \rd x_i \, D_{e/e}(x_i,Q^2) 
  \left\{ \Theta\left( x_i - x_i^0\right) J(x_1,x_2)
  g^\rB(\hat{x},\hat{y},\hat{s}) - g^\rB(x,y,s) \right\} 
\label{bremscross}
\end{equation}
where $\hat{s}  =  x_1 s$, $\hat{x}  =  x x_1 y/(x_1 x_2 + y -1)$,
$\hat{y}  =  y x/x_2 \hat{x}$, $J(x_1,x_2)  =  y/x_1 x_2^2 \hat{y}$,
$x_1^0  =  (1-y)/(1-x y)$ and $x_2^0  =  x y +1-y$, and
\begin{equation}
D_{e/e}(x,Q^2)  = \frac{\alpha}{2\pi}\ln\left(\frac{Q^2}{m_e^2}\right)
\, \frac{1+x^2}{1-x}\ .
\end{equation}
In eq. (\ref{bremscross}) we have supressed the subscript $l$ on 
$x$ and $y$ for clarity. 
Initial-state radiation corresponds to $x_i = x_1$ and $x_2=1$ 
in eq.~(\ref{bremscross}) and vice versa for final-state 
radiation. 
The Compton contribution is given by 
\begin{equation}
  \frac{\rd^2\sigma^{\mathrm{C}}}{\rd x_l \rd y_l} = 
     \int_{\epsilon}^1 \rd z \, f_{\gamma/\re}(z)\, 
  h^{\rC}(x_l,y_l,x_ly_l zs,zs)
\ ,
\label{Compton}
\end{equation}
where 
\begin{eqnarray}
  h^{\rC}(x,y,Q^2,s) &=& \frac{\alpha^3}{x^2(1-y)s}\ Y_+(y)\ 
    \ln\frac{Q^2}{M^2} \int_x^1 \frac{\rd v}{v} 
     \left[ 1 + \left(1 - \frac{x}{v}\right)^2 \right]\nonumber\\
 &\times& F_2(v,Q^2) (1+R(v,Q^2,y))
\ .
\label{hCdef}
\end{eqnarray}
The logarithm $\ln(Q^2/M^2)$ is a result of the absorption
of the collinear singularity from the quark to photon splitting
by renormalizing the photon density in the quark at scale $M$.
We take $M$ here to be the proton mass \cite{Dubna}.

Next we discuss the radiative corrections to the differential
cross section expressed in hadronic variables. Now there are,
in accordance with the KLN theorem \cite{KLN}, in LLA approximation
neither corrections from final state radiation, nor from the
Compton process because these process do not affect the 
kinematic variables. We find that the correction due to 
initial state radiation can simply be expressed as
\begin{eqnarray}
  \frac{\rd^2\sigma^{\rm corr}}{\rd x_h \rd Q_h^2}& = &
 \frac{2 \pi\alpha^2}{x_h Q_h^4}{\Bigg\{}
   F_2(x_h,Q_h^2) \, \int_{z_{min}(x_h,Q_h^2,s)}^1 \rd z\, f_{\gamma/\re}(z)\,
   g_h\left( x_h,Q_h^2,\frac{Q_h^2}{z x_h s}\right) \nonumber \\
 & + &  F_L(x_h,Q_h^2) \, \int_{z_{min}(x_h,Q_h^2,s)}^1 \rd z\, f_{\gamma/\re}(z)\,
   h_h\left( x_h,Q_h^2,\frac{Q_h^2}{z x_h s}\right) {\Bigg\}}
\label{hadcorr}
\end{eqnarray}
where 
\begin{equation}
  g_h(x,Q^2,y) = \frac{\alpha}{\pi}\ \ln \frac{Q^2}{m_e^2}\, 
  \left\{ Y_+(y) \ln (1-y) + y\left(1 - \frac{y}{2}\right) \ln y
  + y\left(1 - \frac{y}{4} \right) \right\}
\end{equation}
and 
\begin{equation}
  h_h(x,Q^2,y) = \frac{\alpha}{\pi}\ \ln \frac{Q^2}{m_e^2}\, 
  \left\{ -y^2 \ln (1-y) + \frac{y^2}{2}\ln y
  - \frac{y}{2}\left(1 - \frac{y}{2} \right) \right\}
\end{equation}
The corrections are large at large $y$ (soft region), and can 
in fact be resummed by simple exponentiation,
with the result 
\begin{eqnarray}
  g_h(x,Q^2,y) & = &  Y_+(y)\ \left\{ \exp \left[ \frac{\alpha}{\pi}\ 
       \left( \ln \frac{Q^2}{m_e^2} - 1 \right) \ln(1-y) \right] 
    - 1 \right\}
\nonumber\\ & & \quad + \frac{\alpha}{\pi}\ \ln \frac{Q^2}{m_e^2}\, 
  \left\{  y\left(1 - \frac{y}{2}\right) \ln y 
  + y\left(1 - \frac{y}{4} \right) \right\}
\label{resummed2}
\end{eqnarray}
and
\begin{eqnarray}
  h_h(x,Q^2,y) & = &  -y^2\ \left\{ \exp \left[ \frac{\alpha}{\pi}\ 
       \left( \ln \frac{Q^2}{m_e^2} - 1 \right) \ln(1-y) \right] 
    - 1 \right\}
\nonumber\\ & & \quad + \frac{\alpha}{\pi}\ \ln \frac{Q^2}{m_e^2}\, 
  \left\{  \left(\frac{y^2}{2}\right) \ln y
  -\frac{y}{2}\left(1 - \frac{y}{2} \right) \right\}
\label{resummedL}
\end{eqnarray}
In the next section we use the formulae listed in the above 
to estimate the size of the corrections.

\section{Results}

Here we study the radiative correction to deep-inelastic electron--photon
scattering numerically. For the results presented below 
we use $\sqrt{s} = 175\,{\rm GeV}$ and $W_{min} = 2\,{\rm GeV}$.
For the parton densities in the photon we use set 1 of \cite{SaS}.
This set has an already low minimum $Q^2$ of $Q_0^2 =0.36\, {\rm GeV}^2$. 
However in radiative events even lower values of $Q^2$ contribute.
We therefore extrapolate below this value by
\begin{eqnarray}
F_2^{\gamma}(x,Q^2<Q_0^2)& = &F_2^{\gamma}(x,Q_0^2)\left(\frac{Q^2}{Q_0^2}\right)^2+ 
  \left(1-\frac{Q^2}{Q_0^2}\right) \frac{Q^2 (1-x)}{112\,{\rm GeV^2}} \nonumber \\
&\times&\left(0.211 (\frac{W^2}{\rm GeV^2})^{0.08} + 
0.297 (\frac{W^2}{\rm GeV^2})^{-0.45}\right), 
\label{extrapol}
\end{eqnarray}
(this expression correctly approaches the $\gamma\gamma$ total cross section
in the small $Q^2$ limit)
which vanishes as $Q^2\rightarrow 0$, as required by gauge invariance.
Furthermore we have neglected $F_L^{\gamma}$, i.e. we
put $R$ in (\ref{Borncross}) 
and (\ref{newBorn}) to zero. We found that its inclusion has
a negligible effect.

\vglue 0.5cm

\setlength{\unitlength}{0.240900pt}
\ifx\plotpoint\undefined\newsavebox{\plotpoint}\fi
\sbox{\plotpoint}{\rule[-0.200pt]{0.400pt}{0.400pt}}%
\begin{picture}(1200,1259)(0,0)
\font\gnuplot=cmr10 at 10pt
\gnuplot
\sbox{\plotpoint}{\rule[-0.200pt]{0.400pt}{0.400pt}}%
\put(176.0,562.0){\rule[-0.200pt]{231.264pt}{0.400pt}}
\put(176.0,113.0){\rule[-0.200pt]{0.400pt}{270.531pt}}
\put(176.0,113.0){\rule[-0.200pt]{4.818pt}{0.400pt}}
\put(154,113){\makebox(0,0)[r]{-0.8}}
\put(1116.0,113.0){\rule[-0.200pt]{4.818pt}{0.400pt}}
\put(176.0,225.0){\rule[-0.200pt]{4.818pt}{0.400pt}}
\put(154,225){\makebox(0,0)[r]{-0.6}}
\put(1116.0,225.0){\rule[-0.200pt]{4.818pt}{0.400pt}}
\put(176.0,338.0){\rule[-0.200pt]{4.818pt}{0.400pt}}
\put(154,338){\makebox(0,0)[r]{-0.4}}
\put(1116.0,338.0){\rule[-0.200pt]{4.818pt}{0.400pt}}
\put(176.0,450.0){\rule[-0.200pt]{4.818pt}{0.400pt}}
\put(154,450){\makebox(0,0)[r]{-0.2}}
\put(1116.0,450.0){\rule[-0.200pt]{4.818pt}{0.400pt}}
\put(176.0,562.0){\rule[-0.200pt]{4.818pt}{0.400pt}}
\put(154,562){\makebox(0,0)[r]{0}}
\put(1116.0,562.0){\rule[-0.200pt]{4.818pt}{0.400pt}}
\put(176.0,675.0){\rule[-0.200pt]{4.818pt}{0.400pt}}
\put(154,675){\makebox(0,0)[r]{0.2}}
\put(1116.0,675.0){\rule[-0.200pt]{4.818pt}{0.400pt}}
\put(176.0,787.0){\rule[-0.200pt]{4.818pt}{0.400pt}}
\put(154,787){\makebox(0,0)[r]{0.4}}
\put(1116.0,787.0){\rule[-0.200pt]{4.818pt}{0.400pt}}
\put(176.0,899.0){\rule[-0.200pt]{4.818pt}{0.400pt}}
\put(154,899){\makebox(0,0)[r]{0.6}}
\put(1116.0,899.0){\rule[-0.200pt]{4.818pt}{0.400pt}}
\put(176.0,1011.0){\rule[-0.200pt]{4.818pt}{0.400pt}}
\put(154,1011){\makebox(0,0)[r]{0.8}}
\put(1116.0,1011.0){\rule[-0.200pt]{4.818pt}{0.400pt}}
\put(176.0,1124.0){\rule[-0.200pt]{4.818pt}{0.400pt}}
\put(154,1124){\makebox(0,0)[r]{1}}
\put(1116.0,1124.0){\rule[-0.200pt]{4.818pt}{0.400pt}}
\put(176.0,1236.0){\rule[-0.200pt]{4.818pt}{0.400pt}}
\put(154,1236){\makebox(0,0)[r]{1.2}}
\put(1116.0,1236.0){\rule[-0.200pt]{4.818pt}{0.400pt}}
\put(176.0,113.0){\rule[-0.200pt]{0.400pt}{4.818pt}}
\put(176,68){\makebox(0,0){0}}
\put(176.0,1216.0){\rule[-0.200pt]{0.400pt}{4.818pt}}
\put(283.0,113.0){\rule[-0.200pt]{0.400pt}{4.818pt}}
\put(283,68){\makebox(0,0){0.1}}
\put(283.0,1216.0){\rule[-0.200pt]{0.400pt}{4.818pt}}
\put(389.0,113.0){\rule[-0.200pt]{0.400pt}{4.818pt}}
\put(389,68){\makebox(0,0){0.2}}
\put(389.0,1216.0){\rule[-0.200pt]{0.400pt}{4.818pt}}
\put(496.0,113.0){\rule[-0.200pt]{0.400pt}{4.818pt}}
\put(496,68){\makebox(0,0){0.3}}
\put(496.0,1216.0){\rule[-0.200pt]{0.400pt}{4.818pt}}
\put(603.0,113.0){\rule[-0.200pt]{0.400pt}{4.818pt}}
\put(603,68){\makebox(0,0){0.4}}
\put(603.0,1216.0){\rule[-0.200pt]{0.400pt}{4.818pt}}
\put(709.0,113.0){\rule[-0.200pt]{0.400pt}{4.818pt}}
\put(709,68){\makebox(0,0){0.5}}
\put(709.0,1216.0){\rule[-0.200pt]{0.400pt}{4.818pt}}
\put(816.0,113.0){\rule[-0.200pt]{0.400pt}{4.818pt}}
\put(816,68){\makebox(0,0){0.6}}
\put(816.0,1216.0){\rule[-0.200pt]{0.400pt}{4.818pt}}
\put(923.0,113.0){\rule[-0.200pt]{0.400pt}{4.818pt}}
\put(923,68){\makebox(0,0){0.7}}
\put(923.0,1216.0){\rule[-0.200pt]{0.400pt}{4.818pt}}
\put(1029.0,113.0){\rule[-0.200pt]{0.400pt}{4.818pt}}
\put(1029,68){\makebox(0,0){0.8}}
\put(1029.0,1216.0){\rule[-0.200pt]{0.400pt}{4.818pt}}
\put(1136.0,113.0){\rule[-0.200pt]{0.400pt}{4.818pt}}
\put(1136,68){\makebox(0,0){0.9}}
\put(1136.0,1216.0){\rule[-0.200pt]{0.400pt}{4.818pt}}
\put(176.0,113.0){\rule[-0.200pt]{231.264pt}{0.400pt}}
\put(1136.0,113.0){\rule[-0.200pt]{0.400pt}{270.531pt}}
\put(176.0,1236.0){\rule[-0.200pt]{231.264pt}{0.400pt}}
\put(656,23){\makebox(0,0){$y_l$}}
\put(176.0,113.0){\rule[-0.200pt]{0.400pt}{270.531pt}}
\put(187,423){\usebox{\plotpoint}}
\multiput(187.58,423.00)(0.498,0.991){81}{\rule{0.120pt}{0.890pt}}
\multiput(186.17,423.00)(42.000,81.152){2}{\rule{0.400pt}{0.445pt}}
\multiput(229.00,506.58)(0.751,0.498){69}{\rule{0.700pt}{0.120pt}}
\multiput(229.00,505.17)(52.547,36.000){2}{\rule{0.350pt}{0.400pt}}
\multiput(283.00,542.58)(1.299,0.498){79}{\rule{1.134pt}{0.120pt}}
\multiput(283.00,541.17)(103.646,41.000){2}{\rule{0.567pt}{0.400pt}}
\multiput(389.00,583.58)(1.928,0.497){53}{\rule{1.629pt}{0.120pt}}
\multiput(389.00,582.17)(103.620,28.000){2}{\rule{0.814pt}{0.400pt}}
\multiput(496.00,611.58)(2.254,0.496){45}{\rule{1.883pt}{0.120pt}}
\multiput(496.00,610.17)(103.091,24.000){2}{\rule{0.942pt}{0.400pt}}
\multiput(603.00,635.58)(1.781,0.497){57}{\rule{1.513pt}{0.120pt}}
\multiput(603.00,634.17)(102.859,30.000){2}{\rule{0.757pt}{0.400pt}}
\multiput(709.00,665.58)(1.632,0.497){63}{\rule{1.397pt}{0.120pt}}
\multiput(709.00,664.17)(104.101,33.000){2}{\rule{0.698pt}{0.400pt}}
\multiput(816.00,698.58)(1.012,0.498){103}{\rule{0.908pt}{0.120pt}}
\multiput(816.00,697.17)(105.116,53.000){2}{\rule{0.454pt}{0.400pt}}
\multiput(923.00,751.58)(0.552,0.498){93}{\rule{0.542pt}{0.120pt}}
\multiput(923.00,750.17)(51.876,48.000){2}{\rule{0.271pt}{0.400pt}}
\multiput(976.58,799.00)(0.498,0.547){103}{\rule{0.120pt}{0.538pt}}
\multiput(975.17,799.00)(53.000,56.884){2}{\rule{0.400pt}{0.269pt}}
\multiput(1029.58,857.00)(0.498,1.096){105}{\rule{0.120pt}{0.974pt}}
\multiput(1028.17,857.00)(54.000,115.978){2}{\rule{0.400pt}{0.487pt}}
\multiput(1083.58,975.00)(0.498,2.085){103}{\rule{0.120pt}{1.760pt}}
\multiput(1082.17,975.00)(53.000,216.346){2}{\rule{0.400pt}{0.880pt}}
\put(187,370){\usebox{\plotpoint}}
\multiput(187.58,370.00)(0.498,0.931){81}{\rule{0.120pt}{0.843pt}}
\multiput(186.17,370.00)(42.000,76.251){2}{\rule{0.400pt}{0.421pt}}
\multiput(229.00,448.58)(0.731,0.498){71}{\rule{0.684pt}{0.120pt}}
\multiput(229.00,447.17)(52.581,37.000){2}{\rule{0.342pt}{0.400pt}}
\multiput(283.00,485.58)(1.331,0.498){77}{\rule{1.160pt}{0.120pt}}
\multiput(283.00,484.17)(103.592,40.000){2}{\rule{0.580pt}{0.400pt}}
\multiput(389.00,525.58)(2.000,0.497){51}{\rule{1.685pt}{0.120pt}}
\multiput(389.00,524.17)(103.502,27.000){2}{\rule{0.843pt}{0.400pt}}
\multiput(496.00,552.58)(2.463,0.496){41}{\rule{2.045pt}{0.120pt}}
\multiput(496.00,551.17)(102.755,22.000){2}{\rule{1.023pt}{0.400pt}}
\multiput(603.00,574.58)(3.174,0.495){31}{\rule{2.594pt}{0.119pt}}
\multiput(603.00,573.17)(100.616,17.000){2}{\rule{1.297pt}{0.400pt}}
\multiput(709.00,591.58)(2.463,0.496){41}{\rule{2.045pt}{0.120pt}}
\multiput(709.00,590.17)(102.755,22.000){2}{\rule{1.023pt}{0.400pt}}
\multiput(816.00,613.58)(1.928,0.497){53}{\rule{1.629pt}{0.120pt}}
\multiput(816.00,612.17)(103.620,28.000){2}{\rule{0.814pt}{0.400pt}}
\multiput(923.00,641.58)(1.490,0.495){33}{\rule{1.278pt}{0.119pt}}
\multiput(923.00,640.17)(50.348,18.000){2}{\rule{0.639pt}{0.400pt}}
\multiput(976.00,659.58)(0.951,0.497){53}{\rule{0.857pt}{0.120pt}}
\multiput(976.00,658.17)(51.221,28.000){2}{\rule{0.429pt}{0.400pt}}
\multiput(1029.00,687.58)(0.587,0.498){89}{\rule{0.570pt}{0.120pt}}
\multiput(1029.00,686.17)(52.818,46.000){2}{\rule{0.285pt}{0.400pt}}
\multiput(1083.58,733.00)(0.498,0.851){103}{\rule{0.120pt}{0.779pt}}
\multiput(1082.17,733.00)(53.000,88.383){2}{\rule{0.400pt}{0.390pt}}
\put(187,343){\usebox{\plotpoint}}
\multiput(187.58,343.00)(0.498,0.919){81}{\rule{0.120pt}{0.833pt}}
\multiput(186.17,343.00)(42.000,75.270){2}{\rule{0.400pt}{0.417pt}}
\multiput(229.00,420.58)(0.731,0.498){71}{\rule{0.684pt}{0.120pt}}
\multiput(229.00,419.17)(52.581,37.000){2}{\rule{0.342pt}{0.400pt}}
\multiput(283.00,457.58)(1.299,0.498){79}{\rule{1.134pt}{0.120pt}}
\multiput(283.00,456.17)(103.646,41.000){2}{\rule{0.567pt}{0.400pt}}
\multiput(389.00,498.58)(2.000,0.497){51}{\rule{1.685pt}{0.120pt}}
\multiput(389.00,497.17)(103.502,27.000){2}{\rule{0.843pt}{0.400pt}}
\multiput(496.00,525.58)(2.714,0.496){37}{\rule{2.240pt}{0.119pt}}
\multiput(496.00,524.17)(102.351,20.000){2}{\rule{1.120pt}{0.400pt}}
\multiput(603.00,545.58)(2.994,0.495){33}{\rule{2.456pt}{0.119pt}}
\multiput(603.00,544.17)(100.903,18.000){2}{\rule{1.228pt}{0.400pt}}
\multiput(709.00,563.58)(3.204,0.495){31}{\rule{2.618pt}{0.119pt}}
\multiput(709.00,562.17)(101.567,17.000){2}{\rule{1.309pt}{0.400pt}}
\multiput(816.00,580.58)(2.714,0.496){37}{\rule{2.240pt}{0.119pt}}
\multiput(816.00,579.17)(102.351,20.000){2}{\rule{1.120pt}{0.400pt}}
\multiput(923.00,600.58)(2.083,0.493){23}{\rule{1.731pt}{0.119pt}}
\multiput(923.00,599.17)(49.408,13.000){2}{\rule{0.865pt}{0.400pt}}
\multiput(976.00,613.58)(1.490,0.495){33}{\rule{1.278pt}{0.119pt}}
\multiput(976.00,612.17)(50.348,18.000){2}{\rule{0.639pt}{0.400pt}}
\multiput(1029.00,631.58)(0.935,0.497){55}{\rule{0.845pt}{0.120pt}}
\multiput(1029.00,630.17)(52.247,29.000){2}{\rule{0.422pt}{0.400pt}}
\multiput(1083.58,660.00)(0.498,0.509){103}{\rule{0.120pt}{0.508pt}}
\multiput(1082.17,660.00)(53.000,52.947){2}{\rule{0.400pt}{0.254pt}}
\put(187,322){\usebox{\plotpoint}}
\multiput(187.58,322.00)(0.498,0.919){81}{\rule{0.120pt}{0.833pt}}
\multiput(186.17,322.00)(42.000,75.270){2}{\rule{0.400pt}{0.417pt}}
\multiput(229.00,399.58)(0.731,0.498){71}{\rule{0.684pt}{0.120pt}}
\multiput(229.00,398.17)(52.581,37.000){2}{\rule{0.342pt}{0.400pt}}
\multiput(283.00,436.58)(1.299,0.498){79}{\rule{1.134pt}{0.120pt}}
\multiput(283.00,435.17)(103.646,41.000){2}{\rule{0.567pt}{0.400pt}}
\multiput(389.00,477.58)(2.000,0.497){51}{\rule{1.685pt}{0.120pt}}
\multiput(389.00,476.17)(103.502,27.000){2}{\rule{0.843pt}{0.400pt}}
\multiput(496.00,504.58)(2.714,0.496){37}{\rule{2.240pt}{0.119pt}}
\multiput(496.00,503.17)(102.351,20.000){2}{\rule{1.120pt}{0.400pt}}
\multiput(603.00,524.58)(3.378,0.494){29}{\rule{2.750pt}{0.119pt}}
\multiput(603.00,523.17)(100.292,16.000){2}{\rule{1.375pt}{0.400pt}}
\multiput(709.00,540.58)(3.204,0.495){31}{\rule{2.618pt}{0.119pt}}
\multiput(709.00,539.17)(101.567,17.000){2}{\rule{1.309pt}{0.400pt}}
\multiput(816.00,557.58)(3.022,0.495){33}{\rule{2.478pt}{0.119pt}}
\multiput(816.00,556.17)(101.857,18.000){2}{\rule{1.239pt}{0.400pt}}
\multiput(923.00,575.58)(2.477,0.492){19}{\rule{2.027pt}{0.118pt}}
\multiput(923.00,574.17)(48.792,11.000){2}{\rule{1.014pt}{0.400pt}}
\multiput(976.00,586.58)(1.929,0.494){25}{\rule{1.614pt}{0.119pt}}
\multiput(976.00,585.17)(49.649,14.000){2}{\rule{0.807pt}{0.400pt}}
\multiput(1029.00,600.58)(1.238,0.496){41}{\rule{1.082pt}{0.120pt}}
\multiput(1029.00,599.17)(51.755,22.000){2}{\rule{0.541pt}{0.400pt}}
\multiput(1083.00,622.58)(0.680,0.498){75}{\rule{0.644pt}{0.120pt}}
\multiput(1083.00,621.17)(51.664,39.000){2}{\rule{0.322pt}{0.400pt}}
\put(187,304){\usebox{\plotpoint}}
\multiput(187.58,304.00)(0.498,0.919){81}{\rule{0.120pt}{0.833pt}}
\multiput(186.17,304.00)(42.000,75.270){2}{\rule{0.400pt}{0.417pt}}
\multiput(229.00,381.58)(0.731,0.498){71}{\rule{0.684pt}{0.120pt}}
\multiput(229.00,380.17)(52.581,37.000){2}{\rule{0.342pt}{0.400pt}}
\multiput(283.00,418.58)(1.299,0.498){79}{\rule{1.134pt}{0.120pt}}
\multiput(283.00,417.17)(103.646,41.000){2}{\rule{0.567pt}{0.400pt}}
\multiput(389.00,459.58)(2.000,0.497){51}{\rule{1.685pt}{0.120pt}}
\multiput(389.00,458.17)(103.502,27.000){2}{\rule{0.843pt}{0.400pt}}
\multiput(496.00,486.58)(2.714,0.496){37}{\rule{2.240pt}{0.119pt}}
\multiput(496.00,485.17)(102.351,20.000){2}{\rule{1.120pt}{0.400pt}}
\multiput(603.00,506.58)(3.174,0.495){31}{\rule{2.594pt}{0.119pt}}
\multiput(603.00,505.17)(100.616,17.000){2}{\rule{1.297pt}{0.400pt}}
\multiput(709.00,523.58)(3.410,0.494){29}{\rule{2.775pt}{0.119pt}}
\multiput(709.00,522.17)(101.240,16.000){2}{\rule{1.388pt}{0.400pt}}
\multiput(816.00,539.58)(3.204,0.495){31}{\rule{2.618pt}{0.119pt}}
\multiput(816.00,538.17)(101.567,17.000){2}{\rule{1.309pt}{0.400pt}}
\multiput(923.00,556.58)(2.737,0.491){17}{\rule{2.220pt}{0.118pt}}
\multiput(923.00,555.17)(48.392,10.000){2}{\rule{1.110pt}{0.400pt}}
\multiput(976.00,566.58)(2.263,0.492){21}{\rule{1.867pt}{0.119pt}}
\multiput(976.00,565.17)(49.126,12.000){2}{\rule{0.933pt}{0.400pt}}
\multiput(1029.00,578.58)(1.519,0.495){33}{\rule{1.300pt}{0.119pt}}
\multiput(1029.00,577.17)(51.302,18.000){2}{\rule{0.650pt}{0.400pt}}
\multiput(1083.00,596.58)(0.887,0.497){57}{\rule{0.807pt}{0.120pt}}
\multiput(1083.00,595.17)(51.326,30.000){2}{\rule{0.403pt}{0.400pt}}
\put(187,287){\usebox{\plotpoint}}
\multiput(187.58,287.00)(0.498,0.919){81}{\rule{0.120pt}{0.833pt}}
\multiput(186.17,287.00)(42.000,75.270){2}{\rule{0.400pt}{0.417pt}}
\multiput(229.00,364.58)(0.731,0.498){71}{\rule{0.684pt}{0.120pt}}
\multiput(229.00,363.17)(52.581,37.000){2}{\rule{0.342pt}{0.400pt}}
\multiput(283.00,401.58)(1.299,0.498){79}{\rule{1.134pt}{0.120pt}}
\multiput(283.00,400.17)(103.646,41.000){2}{\rule{0.567pt}{0.400pt}}
\multiput(389.00,442.58)(2.000,0.497){51}{\rule{1.685pt}{0.120pt}}
\multiput(389.00,441.17)(103.502,27.000){2}{\rule{0.843pt}{0.400pt}}
\multiput(496.00,469.58)(2.714,0.496){37}{\rule{2.240pt}{0.119pt}}
\multiput(496.00,468.17)(102.351,20.000){2}{\rule{1.120pt}{0.400pt}}
\multiput(603.00,489.58)(3.174,0.495){31}{\rule{2.594pt}{0.119pt}}
\multiput(603.00,488.17)(100.616,17.000){2}{\rule{1.297pt}{0.400pt}}
\multiput(709.00,506.58)(3.410,0.494){29}{\rule{2.775pt}{0.119pt}}
\multiput(709.00,505.17)(101.240,16.000){2}{\rule{1.388pt}{0.400pt}}
\multiput(816.00,522.58)(3.204,0.495){31}{\rule{2.618pt}{0.119pt}}
\multiput(816.00,521.17)(101.567,17.000){2}{\rule{1.309pt}{0.400pt}}
\multiput(923.00,539.59)(3.465,0.488){13}{\rule{2.750pt}{0.117pt}}
\multiput(923.00,538.17)(47.292,8.000){2}{\rule{1.375pt}{0.400pt}}
\multiput(976.00,547.58)(2.477,0.492){19}{\rule{2.027pt}{0.118pt}}
\multiput(976.00,546.17)(48.792,11.000){2}{\rule{1.014pt}{0.400pt}}
\multiput(1029.00,558.58)(1.714,0.494){29}{\rule{1.450pt}{0.119pt}}
\multiput(1029.00,557.17)(50.990,16.000){2}{\rule{0.725pt}{0.400pt}}
\multiput(1083.00,574.58)(1.067,0.497){47}{\rule{0.948pt}{0.120pt}}
\multiput(1083.00,573.17)(51.032,25.000){2}{\rule{0.474pt}{0.400pt}}
\put(187,268){\usebox{\plotpoint}}
\multiput(187.58,268.00)(0.498,0.919){81}{\rule{0.120pt}{0.833pt}}
\multiput(186.17,268.00)(42.000,75.270){2}{\rule{0.400pt}{0.417pt}}
\multiput(229.00,345.58)(0.731,0.498){71}{\rule{0.684pt}{0.120pt}}
\multiput(229.00,344.17)(52.581,37.000){2}{\rule{0.342pt}{0.400pt}}
\multiput(283.00,382.58)(1.267,0.498){81}{\rule{1.110pt}{0.120pt}}
\multiput(283.00,381.17)(103.697,42.000){2}{\rule{0.555pt}{0.400pt}}
\multiput(389.00,424.58)(2.078,0.497){49}{\rule{1.746pt}{0.120pt}}
\multiput(389.00,423.17)(103.376,26.000){2}{\rule{0.873pt}{0.400pt}}
\multiput(496.00,450.58)(2.714,0.496){37}{\rule{2.240pt}{0.119pt}}
\multiput(496.00,449.17)(102.351,20.000){2}{\rule{1.120pt}{0.400pt}}
\multiput(603.00,470.58)(2.994,0.495){33}{\rule{2.456pt}{0.119pt}}
\multiput(603.00,469.17)(100.903,18.000){2}{\rule{1.228pt}{0.400pt}}
\multiput(709.00,488.58)(3.644,0.494){27}{\rule{2.953pt}{0.119pt}}
\multiput(709.00,487.17)(100.870,15.000){2}{\rule{1.477pt}{0.400pt}}
\multiput(816.00,503.58)(3.204,0.495){31}{\rule{2.618pt}{0.119pt}}
\multiput(816.00,502.17)(101.567,17.000){2}{\rule{1.309pt}{0.400pt}}
\multiput(923.00,520.59)(3.058,0.489){15}{\rule{2.456pt}{0.118pt}}
\multiput(923.00,519.17)(47.903,9.000){2}{\rule{1.228pt}{0.400pt}}
\multiput(976.00,529.58)(2.737,0.491){17}{\rule{2.220pt}{0.118pt}}
\multiput(976.00,528.17)(48.392,10.000){2}{\rule{1.110pt}{0.400pt}}
\multiput(1029.00,539.58)(2.122,0.493){23}{\rule{1.762pt}{0.119pt}}
\multiput(1029.00,538.17)(50.344,13.000){2}{\rule{0.881pt}{0.400pt}}
\multiput(1083.00,552.58)(1.274,0.496){39}{\rule{1.110pt}{0.119pt}}
\multiput(1083.00,551.17)(50.697,21.000){2}{\rule{0.555pt}{0.400pt}}
\put(187,245){\usebox{\plotpoint}}
\multiput(187.58,245.00)(0.498,0.907){81}{\rule{0.120pt}{0.824pt}}
\multiput(186.17,245.00)(42.000,74.290){2}{\rule{0.400pt}{0.412pt}}
\multiput(229.00,321.58)(0.731,0.498){71}{\rule{0.684pt}{0.120pt}}
\multiput(229.00,320.17)(52.581,37.000){2}{\rule{0.342pt}{0.400pt}}
\multiput(283.00,358.58)(1.267,0.498){81}{\rule{1.110pt}{0.120pt}}
\multiput(283.00,357.17)(103.697,42.000){2}{\rule{0.555pt}{0.400pt}}
\multiput(389.00,400.58)(2.078,0.497){49}{\rule{1.746pt}{0.120pt}}
\multiput(389.00,399.17)(103.376,26.000){2}{\rule{0.873pt}{0.400pt}}
\multiput(496.00,426.58)(2.714,0.496){37}{\rule{2.240pt}{0.119pt}}
\multiput(496.00,425.17)(102.351,20.000){2}{\rule{1.120pt}{0.400pt}}
\multiput(603.00,446.58)(2.994,0.495){33}{\rule{2.456pt}{0.119pt}}
\multiput(603.00,445.17)(100.903,18.000){2}{\rule{1.228pt}{0.400pt}}
\multiput(709.00,464.58)(3.410,0.494){29}{\rule{2.775pt}{0.119pt}}
\multiput(709.00,463.17)(101.240,16.000){2}{\rule{1.388pt}{0.400pt}}
\multiput(816.00,480.58)(3.204,0.495){31}{\rule{2.618pt}{0.119pt}}
\multiput(816.00,479.17)(101.567,17.000){2}{\rule{1.309pt}{0.400pt}}
\multiput(923.00,497.59)(3.465,0.488){13}{\rule{2.750pt}{0.117pt}}
\multiput(923.00,496.17)(47.292,8.000){2}{\rule{1.375pt}{0.400pt}}
\multiput(976.00,505.58)(2.737,0.491){17}{\rule{2.220pt}{0.118pt}}
\multiput(976.00,504.17)(48.392,10.000){2}{\rule{1.110pt}{0.400pt}}
\multiput(1029.00,515.58)(2.122,0.493){23}{\rule{1.762pt}{0.119pt}}
\multiput(1029.00,514.17)(50.344,13.000){2}{\rule{0.881pt}{0.400pt}}
\multiput(1083.00,528.58)(1.580,0.495){31}{\rule{1.347pt}{0.119pt}}
\multiput(1083.00,527.17)(50.204,17.000){2}{\rule{0.674pt}{0.400pt}}
\put(187,213){\usebox{\plotpoint}}
\multiput(187.58,213.00)(0.498,0.907){81}{\rule{0.120pt}{0.824pt}}
\multiput(186.17,213.00)(42.000,74.290){2}{\rule{0.400pt}{0.412pt}}
\multiput(229.00,289.58)(0.751,0.498){69}{\rule{0.700pt}{0.120pt}}
\multiput(229.00,288.17)(52.547,36.000){2}{\rule{0.350pt}{0.400pt}}
\multiput(283.00,325.58)(1.267,0.498){81}{\rule{1.110pt}{0.120pt}}
\multiput(283.00,324.17)(103.697,42.000){2}{\rule{0.555pt}{0.400pt}}
\multiput(389.00,367.58)(2.078,0.497){49}{\rule{1.746pt}{0.120pt}}
\multiput(389.00,366.17)(103.376,26.000){2}{\rule{0.873pt}{0.400pt}}
\multiput(496.00,393.58)(2.714,0.496){37}{\rule{2.240pt}{0.119pt}}
\multiput(496.00,392.17)(102.351,20.000){2}{\rule{1.120pt}{0.400pt}}
\multiput(603.00,413.58)(2.833,0.495){35}{\rule{2.332pt}{0.119pt}}
\multiput(603.00,412.17)(101.161,19.000){2}{\rule{1.166pt}{0.400pt}}
\multiput(709.00,432.58)(3.410,0.494){29}{\rule{2.775pt}{0.119pt}}
\multiput(709.00,431.17)(101.240,16.000){2}{\rule{1.388pt}{0.400pt}}
\multiput(816.00,448.58)(3.204,0.495){31}{\rule{2.618pt}{0.119pt}}
\multiput(816.00,447.17)(101.567,17.000){2}{\rule{1.309pt}{0.400pt}}
\multiput(923.00,465.59)(3.058,0.489){15}{\rule{2.456pt}{0.118pt}}
\multiput(923.00,464.17)(47.903,9.000){2}{\rule{1.228pt}{0.400pt}}
\multiput(976.00,474.59)(3.058,0.489){15}{\rule{2.456pt}{0.118pt}}
\multiput(976.00,473.17)(47.903,9.000){2}{\rule{1.228pt}{0.400pt}}
\multiput(1029.00,483.58)(2.122,0.493){23}{\rule{1.762pt}{0.119pt}}
\multiput(1029.00,482.17)(50.344,13.000){2}{\rule{0.881pt}{0.400pt}}
\multiput(1083.00,496.58)(1.797,0.494){27}{\rule{1.513pt}{0.119pt}}
\multiput(1083.00,495.17)(49.859,15.000){2}{\rule{0.757pt}{0.400pt}}
\put(187,159){\usebox{\plotpoint}}
\multiput(187.58,159.00)(0.498,0.931){81}{\rule{0.120pt}{0.843pt}}
\multiput(186.17,159.00)(42.000,76.251){2}{\rule{0.400pt}{0.421pt}}
\multiput(229.00,237.58)(0.773,0.498){67}{\rule{0.717pt}{0.120pt}}
\multiput(229.00,236.17)(52.512,35.000){2}{\rule{0.359pt}{0.400pt}}
\multiput(283.00,272.58)(1.299,0.498){79}{\rule{1.134pt}{0.120pt}}
\multiput(283.00,271.17)(103.646,41.000){2}{\rule{0.567pt}{0.400pt}}
\multiput(389.00,313.58)(2.078,0.497){49}{\rule{1.746pt}{0.120pt}}
\multiput(389.00,312.17)(103.376,26.000){2}{\rule{0.873pt}{0.400pt}}
\multiput(496.00,339.58)(2.714,0.496){37}{\rule{2.240pt}{0.119pt}}
\multiput(496.00,338.17)(102.351,20.000){2}{\rule{1.120pt}{0.400pt}}
\multiput(603.00,359.58)(2.833,0.495){35}{\rule{2.332pt}{0.119pt}}
\multiput(603.00,358.17)(101.161,19.000){2}{\rule{1.166pt}{0.400pt}}
\multiput(709.00,378.58)(3.410,0.494){29}{\rule{2.775pt}{0.119pt}}
\multiput(709.00,377.17)(101.240,16.000){2}{\rule{1.388pt}{0.400pt}}
\multiput(816.00,394.58)(3.022,0.495){33}{\rule{2.478pt}{0.119pt}}
\multiput(816.00,393.17)(101.857,18.000){2}{\rule{1.239pt}{0.400pt}}
\multiput(923.00,412.59)(3.058,0.489){15}{\rule{2.456pt}{0.118pt}}
\multiput(923.00,411.17)(47.903,9.000){2}{\rule{1.228pt}{0.400pt}}
\multiput(976.00,421.58)(2.737,0.491){17}{\rule{2.220pt}{0.118pt}}
\multiput(976.00,420.17)(48.392,10.000){2}{\rule{1.110pt}{0.400pt}}
\multiput(1029.00,431.58)(2.122,0.493){23}{\rule{1.762pt}{0.119pt}}
\multiput(1029.00,430.17)(50.344,13.000){2}{\rule{0.881pt}{0.400pt}}
\multiput(1083.00,444.58)(1.929,0.494){25}{\rule{1.614pt}{0.119pt}}
\multiput(1083.00,443.17)(49.649,14.000){2}{\rule{0.807pt}{0.400pt}}
\end{picture}

{\tenrm\baselineskip=12pt 
\noindent Figure 2: The ratio $\delta(x_l,y_l)$ (\ref{delxy}) vs.
$y_l$ for various $x_l$ values. Top curve is for
$x_l = 0.01$, the next for $x_l=0.1$. Each subsequent
curve represents an increase of $x_l$ by $0.1$.}

\vglue 0.2cm

For comparison with the $\re$p case  we now show in Fig.2 the correction
$\delta(x_l,y_l)$, using (\ref{nsigBorn}), (\ref{bremscross}) and 
(\ref{Compton}),
 defined by
\begin{equation}
\frac{\rd^2\sigma^{\mathrm{Br}}}{\rd x_l \rd y_l} 
+  \frac{\rd^2\sigma^{\mathrm{C}}}{\rd x_l \rd y_l}=
\frac{\rd^2\sigma^\rB}{\rd x_l \rd y_l}\,\,\delta(x_l,y_l)\ .
\label{delxy}
\end{equation}
Note that we use here the leptonic variables to conform with
the $\re$p case. 
A closer examination reveals
that initial and final state radiation are similar in order
of magnitude throughout most of the $x,y$ region, 
whereas the Compton contributions is 
appreciable only in the small and medium  $x$, large $y$ region.
Also we note that if one freezes $F_2$ at $F_2(Q_0^2)$ for
$Q^2<Q^2_0$, instead of extrapolating as in (\ref{extrapol}),
we find that only the $x_l=0.01$ curve changes significantly.
It decreases at small and medium $y_l$ by up to 50\%.
Note that inclusion of final state radiation implies a perfect
measurement of the energy and momentum of the tagged lepton,
even in the presence of collinear radiation.

We see in Fig.2 that radiative effects can in principle be
large, around 40\% for medium $x$ and small $y$.

Next we show in Fig.3 the correction factor 
$\delta(x_h,Q^2_h)$, defined by
\begin{equation}
\frac{\rd^2\sigma^{\mathrm{corr}}}{\rd x_h \rd Q^2_h} =
\frac{\rd^2\sigma^\rB}{\rd x_h \rd Q^2_h} \delta(x_h,Q^2_h)\ .
\label{delxQ2}, 
\end{equation}
in terms of hadronic variables, 
as a function of $x_h$ for various choices of
$Q_h^2$, cf. (\ref{hadcorr}). Here only initial state lepton
bremsstrahlung is taken into account, because the scattered lepton
is not used in constructing the kinematic variables.

\vglue 0.5cm
%

\setlength{\unitlength}{0.240900pt}
\ifx\plotpoint\undefined\newsavebox{\plotpoint}\fi
\sbox{\plotpoint}{\rule[-0.200pt]{0.400pt}{0.400pt}}%
\begin{picture}(1500,900)(0,0)
\font\gnuplot=cmr10 at 10pt
\gnuplot
\sbox{\plotpoint}{\rule[-0.200pt]{0.400pt}{0.400pt}}%
\put(176.0,113.0){\rule[-0.200pt]{0.400pt}{184.048pt}}
\put(176.0,113.0){\rule[-0.200pt]{4.818pt}{0.400pt}}
\put(154,113){\makebox(0,0)[r]{-0.4}}
\put(1416.0,113.0){\rule[-0.200pt]{4.818pt}{0.400pt}}
\put(176.0,209.0){\rule[-0.200pt]{4.818pt}{0.400pt}}
\put(154,209){\makebox(0,0)[r]{-0.35}}
\put(1416.0,209.0){\rule[-0.200pt]{4.818pt}{0.400pt}}
\put(176.0,304.0){\rule[-0.200pt]{4.818pt}{0.400pt}}
\put(154,304){\makebox(0,0)[r]{-0.3}}
\put(1416.0,304.0){\rule[-0.200pt]{4.818pt}{0.400pt}}
\put(176.0,399.0){\rule[-0.200pt]{4.818pt}{0.400pt}}
\put(154,399){\makebox(0,0)[r]{-0.25}}
\put(1416.0,399.0){\rule[-0.200pt]{4.818pt}{0.400pt}}
\put(176.0,495.0){\rule[-0.200pt]{4.818pt}{0.400pt}}
\put(154,495){\makebox(0,0)[r]{-0.2}}
\put(1416.0,495.0){\rule[-0.200pt]{4.818pt}{0.400pt}}
\put(176.0,590.0){\rule[-0.200pt]{4.818pt}{0.400pt}}
\put(154,590){\makebox(0,0)[r]{-0.15}}
\put(1416.0,590.0){\rule[-0.200pt]{4.818pt}{0.400pt}}
\put(176.0,686.0){\rule[-0.200pt]{4.818pt}{0.400pt}}
\put(154,686){\makebox(0,0)[r]{-0.1}}
\put(1416.0,686.0){\rule[-0.200pt]{4.818pt}{0.400pt}}
\put(176.0,782.0){\rule[-0.200pt]{4.818pt}{0.400pt}}
\put(154,782){\makebox(0,0)[r]{-0.05}}
\put(1416.0,782.0){\rule[-0.200pt]{4.818pt}{0.400pt}}
\put(176.0,877.0){\rule[-0.200pt]{4.818pt}{0.400pt}}
\put(154,877){\makebox(0,0)[r]{0}}
\put(1416.0,877.0){\rule[-0.200pt]{4.818pt}{0.400pt}}
\put(176.0,113.0){\rule[-0.200pt]{0.400pt}{4.818pt}}
\put(176,68){\makebox(0,0){0}}
\put(176.0,857.0){\rule[-0.200pt]{0.400pt}{4.818pt}}
\put(302.0,113.0){\rule[-0.200pt]{0.400pt}{4.818pt}}
\put(302,68){\makebox(0,0){0.1}}
\put(302.0,857.0){\rule[-0.200pt]{0.400pt}{4.818pt}}
\put(428.0,113.0){\rule[-0.200pt]{0.400pt}{4.818pt}}
\put(428,68){\makebox(0,0){0.2}}
\put(428.0,857.0){\rule[-0.200pt]{0.400pt}{4.818pt}}
\put(554.0,113.0){\rule[-0.200pt]{0.400pt}{4.818pt}}
\put(554,68){\makebox(0,0){0.3}}
\put(554.0,857.0){\rule[-0.200pt]{0.400pt}{4.818pt}}
\put(680.0,113.0){\rule[-0.200pt]{0.400pt}{4.818pt}}
\put(680,68){\makebox(0,0){0.4}}
\put(680.0,857.0){\rule[-0.200pt]{0.400pt}{4.818pt}}
\put(806.0,113.0){\rule[-0.200pt]{0.400pt}{4.818pt}}
\put(806,68){\makebox(0,0){0.5}}
\put(806.0,857.0){\rule[-0.200pt]{0.400pt}{4.818pt}}
\put(932.0,113.0){\rule[-0.200pt]{0.400pt}{4.818pt}}
\put(932,68){\makebox(0,0){0.6}}
\put(932.0,857.0){\rule[-0.200pt]{0.400pt}{4.818pt}}
\put(1058.0,113.0){\rule[-0.200pt]{0.400pt}{4.818pt}}
\put(1058,68){\makebox(0,0){0.7}}
\put(1058.0,857.0){\rule[-0.200pt]{0.400pt}{4.818pt}}
\put(1184.0,113.0){\rule[-0.200pt]{0.400pt}{4.818pt}}
\put(1184,68){\makebox(0,0){0.8}}
\put(1184.0,857.0){\rule[-0.200pt]{0.400pt}{4.818pt}}
\put(1310.0,113.0){\rule[-0.200pt]{0.400pt}{4.818pt}}
\put(1310,68){\makebox(0,0){0.9}}
\put(1310.0,857.0){\rule[-0.200pt]{0.400pt}{4.818pt}}
\put(1436.0,113.0){\rule[-0.200pt]{0.400pt}{4.818pt}}
\put(1436,68){\makebox(0,0){1}}
\put(1436.0,857.0){\rule[-0.200pt]{0.400pt}{4.818pt}}
\put(176.0,113.0){\rule[-0.200pt]{303.534pt}{0.400pt}}
\put(1436.0,113.0){\rule[-0.200pt]{0.400pt}{184.048pt}}
\put(176.0,877.0){\rule[-0.200pt]{303.534pt}{0.400pt}}
\put(806,23){\makebox(0,0){$x_h$}}
\put(176.0,113.0){\rule[-0.200pt]{0.400pt}{184.048pt}}
\put(189,698){\usebox{\plotpoint}}
\multiput(189.00,698.58)(0.625,0.498){77}{\rule{0.600pt}{0.120pt}}
\multiput(189.00,697.17)(48.755,40.000){2}{\rule{0.300pt}{0.400pt}}
\multiput(239.00,738.58)(2.479,0.493){23}{\rule{2.038pt}{0.119pt}}
\multiput(239.00,737.17)(58.769,13.000){2}{\rule{1.019pt}{0.400pt}}
\multiput(302.00,751.59)(6.944,0.477){7}{\rule{5.140pt}{0.115pt}}
\multiput(302.00,750.17)(52.332,5.000){2}{\rule{2.570pt}{0.400pt}}
\multiput(365.00,756.60)(9.108,0.468){5}{\rule{6.400pt}{0.113pt}}
\multiput(365.00,755.17)(49.716,4.000){2}{\rule{3.200pt}{0.400pt}}
\put(428,760.17){\rule{12.700pt}{0.400pt}}
\multiput(428.00,759.17)(36.641,2.000){2}{\rule{6.350pt}{0.400pt}}
\put(491,762.17){\rule{12.700pt}{0.400pt}}
\multiput(491.00,761.17)(36.641,2.000){2}{\rule{6.350pt}{0.400pt}}
\multiput(554.00,764.61)(13.858,0.447){3}{\rule{8.500pt}{0.108pt}}
\multiput(554.00,763.17)(45.358,3.000){2}{\rule{4.250pt}{0.400pt}}
\put(617,765.67){\rule{15.177pt}{0.400pt}}
\multiput(617.00,766.17)(31.500,-1.000){2}{\rule{7.588pt}{0.400pt}}
\multiput(680.00,766.59)(6.944,0.477){7}{\rule{5.140pt}{0.115pt}}
\multiput(680.00,765.17)(52.332,5.000){2}{\rule{2.570pt}{0.400pt}}
\put(743,771.17){\rule{12.700pt}{0.400pt}}
\multiput(743.00,770.17)(36.641,2.000){2}{\rule{6.350pt}{0.400pt}}
\put(806,772.67){\rule{15.177pt}{0.400pt}}
\multiput(806.00,772.17)(31.500,1.000){2}{\rule{7.588pt}{0.400pt}}
\put(932,773.67){\rule{15.177pt}{0.400pt}}
\multiput(932.00,773.17)(31.500,1.000){2}{\rule{7.588pt}{0.400pt}}
\multiput(995.00,773.95)(13.858,-0.447){3}{\rule{8.500pt}{0.108pt}}
\multiput(995.00,774.17)(45.358,-3.000){2}{\rule{4.250pt}{0.400pt}}
\multiput(1058.00,772.61)(13.858,0.447){3}{\rule{8.500pt}{0.108pt}}
\multiput(1058.00,771.17)(45.358,3.000){2}{\rule{4.250pt}{0.400pt}}
\put(1121,775.17){\rule{12.700pt}{0.400pt}}
\multiput(1121.00,774.17)(36.641,2.000){2}{\rule{6.350pt}{0.400pt}}
\multiput(1184.00,775.95)(13.858,-0.447){3}{\rule{8.500pt}{0.108pt}}
\multiput(1184.00,776.17)(45.358,-3.000){2}{\rule{4.250pt}{0.400pt}}
\multiput(1247.00,774.61)(13.858,0.447){3}{\rule{8.500pt}{0.108pt}}
\multiput(1247.00,773.17)(45.358,3.000){2}{\rule{4.250pt}{0.400pt}}
\put(1310,777.17){\rule{12.700pt}{0.400pt}}
\multiput(1310.00,776.17)(36.641,2.000){2}{\rule{6.350pt}{0.400pt}}
\put(869.0,774.0){\rule[-0.200pt]{15.177pt}{0.400pt}}
\put(189,543){\usebox{\plotpoint}}
\multiput(189.58,543.00)(0.498,0.993){97}{\rule{0.120pt}{0.892pt}}
\multiput(188.17,543.00)(50.000,97.149){2}{\rule{0.400pt}{0.446pt}}
\multiput(239.00,642.58)(1.132,0.497){53}{\rule{1.000pt}{0.120pt}}
\multiput(239.00,641.17)(60.924,28.000){2}{\rule{0.500pt}{0.400pt}}
\multiput(302.00,670.58)(2.139,0.494){27}{\rule{1.780pt}{0.119pt}}
\multiput(302.00,669.17)(59.306,15.000){2}{\rule{0.890pt}{0.400pt}}
\multiput(365.00,685.59)(3.640,0.489){15}{\rule{2.900pt}{0.118pt}}
\multiput(365.00,684.17)(56.981,9.000){2}{\rule{1.450pt}{0.400pt}}
\multiput(428.00,694.59)(6.944,0.477){7}{\rule{5.140pt}{0.115pt}}
\multiput(428.00,693.17)(52.332,5.000){2}{\rule{2.570pt}{0.400pt}}
\multiput(491.00,699.59)(5.644,0.482){9}{\rule{4.300pt}{0.116pt}}
\multiput(491.00,698.17)(54.075,6.000){2}{\rule{2.150pt}{0.400pt}}
\multiput(554.00,705.59)(5.644,0.482){9}{\rule{4.300pt}{0.116pt}}
\multiput(554.00,704.17)(54.075,6.000){2}{\rule{2.150pt}{0.400pt}}
\multiput(680.00,711.60)(9.108,0.468){5}{\rule{6.400pt}{0.113pt}}
\multiput(680.00,710.17)(49.716,4.000){2}{\rule{3.200pt}{0.400pt}}
\put(617.0,711.0){\rule[-0.200pt]{15.177pt}{0.400pt}}
\multiput(806.00,715.60)(9.108,0.468){5}{\rule{6.400pt}{0.113pt}}
\multiput(806.00,714.17)(49.716,4.000){2}{\rule{3.200pt}{0.400pt}}
\multiput(869.00,719.61)(13.858,0.447){3}{\rule{8.500pt}{0.108pt}}
\multiput(869.00,718.17)(45.358,3.000){2}{\rule{4.250pt}{0.400pt}}
\put(932,722.17){\rule{12.700pt}{0.400pt}}
\multiput(932.00,721.17)(36.641,2.000){2}{\rule{6.350pt}{0.400pt}}
\put(995,722.67){\rule{15.177pt}{0.400pt}}
\multiput(995.00,723.17)(31.500,-1.000){2}{\rule{7.588pt}{0.400pt}}
\multiput(1058.00,723.61)(13.858,0.447){3}{\rule{8.500pt}{0.108pt}}
\multiput(1058.00,722.17)(45.358,3.000){2}{\rule{4.250pt}{0.400pt}}
\put(1121,725.67){\rule{15.177pt}{0.400pt}}
\multiput(1121.00,725.17)(31.500,1.000){2}{\rule{7.588pt}{0.400pt}}
\put(743.0,715.0){\rule[-0.200pt]{15.177pt}{0.400pt}}
\multiput(1247.00,727.61)(13.858,0.447){3}{\rule{8.500pt}{0.108pt}}
\multiput(1247.00,726.17)(45.358,3.000){2}{\rule{4.250pt}{0.400pt}}
\put(1184.0,727.0){\rule[-0.200pt]{15.177pt}{0.400pt}}
\multiput(194.58,113.00)(0.498,3.467){87}{\rule{0.120pt}{2.856pt}}
\multiput(193.17,113.00)(45.000,304.073){2}{\rule{0.400pt}{1.428pt}}
\multiput(239.58,423.00)(0.499,0.627){123}{\rule{0.120pt}{0.602pt}}
\multiput(238.17,423.00)(63.000,77.751){2}{\rule{0.400pt}{0.301pt}}
\multiput(302.00,502.58)(0.930,0.498){65}{\rule{0.841pt}{0.120pt}}
\multiput(302.00,501.17)(61.254,34.000){2}{\rule{0.421pt}{0.400pt}}
\multiput(365.00,536.58)(1.593,0.496){37}{\rule{1.360pt}{0.119pt}}
\multiput(365.00,535.17)(60.177,20.000){2}{\rule{0.680pt}{0.400pt}}
\multiput(428.00,556.58)(1.881,0.495){31}{\rule{1.582pt}{0.119pt}}
\multiput(428.00,555.17)(59.716,17.000){2}{\rule{0.791pt}{0.400pt}}
\multiput(491.00,573.58)(2.693,0.492){21}{\rule{2.200pt}{0.119pt}}
\multiput(491.00,572.17)(58.434,12.000){2}{\rule{1.100pt}{0.400pt}}
\multiput(554.00,585.59)(4.126,0.488){13}{\rule{3.250pt}{0.117pt}}
\multiput(554.00,584.17)(56.254,8.000){2}{\rule{1.625pt}{0.400pt}}
\multiput(617.00,593.59)(3.640,0.489){15}{\rule{2.900pt}{0.118pt}}
\multiput(617.00,592.17)(56.981,9.000){2}{\rule{1.450pt}{0.400pt}}
\multiput(680.00,602.59)(5.644,0.482){9}{\rule{4.300pt}{0.116pt}}
\multiput(680.00,601.17)(54.075,6.000){2}{\rule{2.150pt}{0.400pt}}
\multiput(743.00,608.59)(6.944,0.477){7}{\rule{5.140pt}{0.115pt}}
\multiput(743.00,607.17)(52.332,5.000){2}{\rule{2.570pt}{0.400pt}}
\multiput(806.00,613.59)(6.944,0.477){7}{\rule{5.140pt}{0.115pt}}
\multiput(806.00,612.17)(52.332,5.000){2}{\rule{2.570pt}{0.400pt}}
\multiput(869.00,618.60)(9.108,0.468){5}{\rule{6.400pt}{0.113pt}}
\multiput(869.00,617.17)(49.716,4.000){2}{\rule{3.200pt}{0.400pt}}
\multiput(932.00,622.59)(5.644,0.482){9}{\rule{4.300pt}{0.116pt}}
\multiput(932.00,621.17)(54.075,6.000){2}{\rule{2.150pt}{0.400pt}}
\put(995,627.67){\rule{15.177pt}{0.400pt}}
\multiput(995.00,627.17)(31.500,1.000){2}{\rule{7.588pt}{0.400pt}}
\put(1058,629.17){\rule{12.700pt}{0.400pt}}
\multiput(1058.00,628.17)(36.641,2.000){2}{\rule{6.350pt}{0.400pt}}
\multiput(1121.00,631.59)(6.944,0.477){7}{\rule{5.140pt}{0.115pt}}
\multiput(1121.00,630.17)(52.332,5.000){2}{\rule{2.570pt}{0.400pt}}
\multiput(1184.00,636.61)(13.858,0.447){3}{\rule{8.500pt}{0.108pt}}
\multiput(1184.00,635.17)(45.358,3.000){2}{\rule{4.250pt}{0.400pt}}
\put(1247,638.67){\rule{15.177pt}{0.400pt}}
\multiput(1247.00,638.17)(31.500,1.000){2}{\rule{7.588pt}{0.400pt}}
\put(1310,640.17){\rule{12.700pt}{0.400pt}}
\multiput(1310.00,639.17)(36.641,2.000){2}{\rule{6.350pt}{0.400pt}}
\put(1310.0,730.0){\rule[-0.200pt]{15.177pt}{0.400pt}}
\end{picture}

{\tenrm\baselineskip=12pt 
\noindent Figure 3: The ratio $\delta(x_h,Q^2_h)$  vs.
$x_h$ for three $Q^2_h$ values. Top curve: $Q^2_h = 1 \, {\rm GeV^2}$.
Middle curve: $Q^2_h = 10\, {\rm GeV^2}$. Lower curve:
$Q^2_h = 100\, {\rm GeV^2}$.}

\vglue 0.5cm

We see that the corrections are sizable only for large $Q^2$.
Using however the resummed version of (\ref{resummed2}) and 
(\ref{resummedL}), we find
that the corrections are reduced by about an order
of magnitude.

Thus we conclude that the size of the radiative corrections depends
significantly on the set of variables chosen. These corrections
can in principle be quite large. As noted before, in practice
mixed variables are used. In view of the results obtained in this paper, 
we think a more careful study, involving Monte Carlo simulation of 
the full final state, is warranted.

\end{document}